# Exploring the Role of Perceived Range Anxiety in Adoption Behavior of Plug-in Electric Vehicles

Fatemeh Nazari[1,a], Abolfazl (Kouros) Mohammadian [b]

[a] *Department of Civil Engineering, The University of Texas Rio Grande Valley, Edinburg, TX 78539, United States*

[b] *Department of Civil, Environmental, and Materials Engineering, University of Illinois at Chicago, Chicago, IL 60607, United States*

**Abstract:** A sustainable solution to negative externalities imposed by road transportation is replacing internal combustion vehicles with electric vehicles (EVs), especially plug-in EV (PEV) encompassing plug-in hybrid EV (PHEV) and battery EV (BEV). However, EV market share is still low and is forecast to remain low and uncertain. This shows a research need for an in-depth understanding of EV adoption behavior with a focus on one of the main barriers to the mass EV adoption, which is the limited electric driving range. The present study extends the existing literature in two directions; First, the influence of the *psychological aspect* of driving range, which is referred to as "range anxiety", is explored on EV adoption behavior by presenting a nested logit (NL) model with a latent construct. Second, the two-level NL model captures individuals' decision on EV adoption behavior distinguished by *vehicle transaction type* and EV type, where the upper level yields the vehicle transaction type selected from the set of alternatives including no-transaction, sell, trade, and add. The fuel type of the vehicles decided to be acquired, either as traded-for or added vehicles, is simultaneously determined at the lower level from a set including conventional vehicle, hybrid EV, PHEV, and BEV. The model is empirically estimated using a stated preferences dataset collected in the State of California. A notable finding is that anxiety about driving range influences the preference for BEV, especially as an added than traded-for vehicle, but not the decision on PHEV adoption.

[1] Corresponding author: Fatemeh Nazari, fatemeh.nazari@utrgv.edu

## 1. Introduction

Nations around the globe are struggling with negative mobility externalities, such as traffic congestion, air pollution, accidents, noise, and costs of energy dependencies, caused in part by the prevalence of internal combustion vehicles.[1–3] This persistent problem can be mitigated by substituting conventional gasoline and diesel vehicles (CVs) with electric vehicles (EVs), which encompass hybrid EV (HEV), plug-in HEV (PHEV), and battery EV (BEV). To achieve this sustainability goal, various EV types need to be accepted and then adopted by the public on a large scale. However, EV market share is still low in the U.S. and worldwide despite its growth and is predicted to remain low and uncertain in the short to midterm.[4–6] The factors impeding the extensive adoption of EVs, especially PHEV and BEV which are collectively referred to as plug-in EV (PEV), are high purchase price, long recharging time, lack of charging infrastructure, and limited electric driving range.[7–9] Among them, the driving range factor is found to contribute substantially.[10–14]

Driving range is *technically* improving in two ways; First, the advances in battery technology have led to, for instance, the advent of affordable BEVs with ranges above 200 miles in the 2020 vehicle market.[15,16] Second, numerous policy decisions are made to improve public charging infrastructure such as the $7.5 billion investment in charging infrastructure by the U.S. government under the Bipartisan Infrastructure Law.[5,17] With the technical improvements in the battery technology and the charging infrastructure, the driving range of the existing PEVs are shown to satisfy the travel needs of a typical driver.[18–20] Nevertheless, the driving range factor still persists as a major barrier to PEV adoption.[21] This implies that one needs to look into the non-technical aspect of driving range, namely, the *psychological* aspect,[22,13] which is referred to as "range anxiety." According to Ref. 23, range anxiety reflects the psychological anxiety caused by an irrational fear response unjustifiable by the driving range needed in reality.

The economic-based studies examining the EV adoption behavior (summarized in Table 1) mostly lay greater stress on the technical rather than the psychological aspect of driving range. To the best of our knowledge, one exception is the study by Ref. 24 who investigate the relevant impact of range concern defined as an explanatory binary variable. To fill this gap, the first contribution of the present study is understanding whether and how range anxiety shapes EV adoption behavior. To do so, a latent subjective construct is built on the individuals' perceived anxiety about electric driving range. By estimating a confirmatory factor analysis (CFA), the latent construct is related to the corresponding indicators, which measure the individuals' responses to five questions on their opinions about perception of range anxiety. The estimated latent construct is subsequently plugged into a two-level nested logit (NL) model. The nesting structure of the NL model is the second contribution of this study, which *simultaneously* determines the individuals' adoption behavior of various EV types and their decision on vehicle transaction type. In particular, the first level yields vehicle transaction type selected from the set of four alternatives, including no-transaction, sell, trade, and add. The fuel type of the vehicles decided to be acquired, either as traded-for or added vehicles, is jointly determined at the second level of the NL model. The fuel type options include CV, HEV, PHEV, and BEV. This joint structure of the model is necessary for an in-depth understanding of the adoption behavior of various EV types by distinctly characterizing those who are interested in trading an existing vehicle for an EV type as well as individuals inclined towards adding an EV type to the existing household vehicle(s).

The remainder of the paper is organized as follows. The relevant literature is reviewed in section 2 followed by the modeling framework presented in section 3. The framework is then empirically estimated using a dataset statistically described in section 4 with findings analyzed in section 5. Section 6 delivers an overview of the study and suggestions for the future research.

**Table 1. A summary of studies applying discrete choice models to explore EV adoption behavior**

| | **Methodology** | | | | | | **Driving range factor** | |
|---|---|---|---|---|---|---|---|---|
| | | **Fuel type alternative** | | | | | | |
| **Study, location** | **Choice model** | **CV** | **HEV** | **PHEV** | **BEV** | **Joint vehicle decisions *** | **Technical** | **Psychological** |
| Beggs et al. (1981),[25] U.S. | Ordered logit | √ | | | √ | | √ | |
| Brownstone et al. (1996),[26] U.S. | Nested logit | √ | | | √ | √ | √ | |
| Maness & Cirillo (2012),[27] U.S. | Mixed logit | √ | √ | | √ | √ | √ | |
| Hoen & Koetse (2014),[28] Netherlands | Mixed logit | √ | | | √ | | √ | |
| Hackbarth & Madlener (2016),[11] Germany | Latent class | √ | √ | √ | √ | | √ | |
| Smith et al. (2017),[29] Australia | Best-worst | √ | | | √ | | √ | |
| Cirillo et al. (2017),[30] U.S. | Mixed logit | √ | √ | | √ | √ | √ | |
| Liao et al. (2018),[31] Netherlands | Latent class | √ | | √ | √ | | √ | |
| Wolbertus et al. (2018),[32] Netherlands | Mixed logit | √ | | √ | √ | | √ | |
| Kwon et al. (2018),[33] Korea | Multinomial logit | | | | √ | | √ | |
| Lane et al. (2018),[24] U.S. | Multinomial probit & binary logit | √ | √ | √ | √ | | | √ |
| Leard (2018),[34] U.S. | Mixed logit | √ | √ | √ | √ | | √ | |
| Liao et al. (2019),[35] Netherlands | Mixed logit & hybrid choice | √ | | √ | √ | | √ | |
| Qian et al. (2019),[36] China | Mixed logit | √ | | √ | √ | | √ | |
| Ghasri et al. (2019),[37] Australia | Hybrid choice | | | | √ | | √ | |

| Study, location | Choice model | Methodology | | | | | Driving range factor | |
|---|---|---|---|---|---|---|---|---|
| | | Fuel type alternative | | | | Joint vehicle decisions * | Technical | Psychological |
| | | CV | HEV | PHEV | BEV | | | |
| Li et al. (2020),[38] China | Mixed logit & latent class | √ | | | √ | | √ | |
| Guerra & Daziano (2020),[39] U.S. | Mixed logit &latent class | √ | | | √ | | √ | |
| Bansal et al. (2021),[40] India | Hybrid choice | √ | | | √ | | √ | |
| Khan et al. (2021),[41] Canada | Latent class | √ | √ | √ | √ | | √ | |
| Jia & Chen (2021),[42] U.S. | Mixed logit | √ | √ | √ | √ | | √ | |
| Munshi et al. (2022),[21] India | Binary logit | | | | √ | | √ | |
| Jordan Hamed et al. (2023),[9] Jordan | Random parameters ordered probit | | | | √ | | √ | |
| **Present study, U.S.** | **Nested logit with a latent construct** | √ | √ | √ | √ | √ | | √ |

Note: * = such as vehicle fuel type and vehicle transaction; CV = conventional gasoline and diesel vehicle; HEV = hybrid electric vehicle; PHEV = plug-in hybrid electric vehicle; BEV = battery electric vehicle; PEV = PHEV or BEV.

# 2. Literature Review

## 2.1. Electric Driving Range: Technical Aspect

A great body of the relevant literature considers the technical aspect of driving range as a PEV attribute to analyze the corresponding impact on the behavioral (intention) to accept or adopt PEV at disaggregate-level (e.g., individual). Such studies mostly develop modeling frameworks based on psychology (such as Theory of Planned Behavior, Theory of Reasoned Action, and Diffusion of Innovation) or economics (such as discrete choice models). The former associates an individual's behavioral intention to accept, adopt, or use EV merely with his/her unobserved (latent) subjective attitude, perception, emotion, and symbolism labeled as "taste heterogeneity" (see a review in Ref. 43 and examples of empirical studies in Ref. 44, Ref. 45, and Ref. 46). The latter, which is employed in the present study, is capable of explaining the choice process by both latent and observed heterogeneity in addition to providing a suitable framework to evaluate policy effectiveness (see Ref. 47 for a review on the empirical studies). The observed heterogeneity stems from the individuals' observed characteristics such as socio-economic attributes.

The modeling frameworks used in the economic-based approach, which are typically a discrete choice model, are usually empirically estimated on datasets collected through designing and conducing stated preferences (SP) experiments. SP datasets capture the individuals' "hypothetical" choice from a set of given alternatives rather than their "actual" choice from a set of available alternatives, where the second one can be measured by collecting revealed preferences (RP) datasets. This assumption is acceptable in the context of EV adoption modeling given that the corresponding market share is still small. For instance, in the U.S. HEV was initially introduced almost two decades ago in 2000 and PEV entered the market in 2010,[6] which collectively constitute less than 2% of the market share. To our best knowledge, there are limited choice-based studies on EV adoption behavior using RP datasets collected in the U.S. The examples are Ref. 48, Ref. 49, Ref. 50, and Ref. 51, yet none of them explore the role of driving range on EV adoption behavior.

A number of empirical studies close to the present study is summarized in Table 1 with details on the study location, method, and driving range aspects. A pioneering study is conducted by Ref. 25 who highlight the negative impact of vehicle driving range on the demand for BEV by estimating an ordered logit model on a dataset measuring individuals' responses to rank ordering 16 vehicle types in nine U.S. cities. Another example of the earlier studies is the research by Ref. 26 who analyze the joint decision of vehicle transaction and fuel type choice of one- and two-vehicle households. By estimating a nested logit model on a dataset collected in the State of California, U.S., the authors report vehicle driving range as an important factor in the households' decision on EV acquisition.

Later on, Ref. 28 investigate the preferences of the Dutch population for alternative fuel vehicles (including CV, BEV, and fuel-cell vehicle) by estimating a mixed logit model. They find that the considerable negative tendency towards BEV is caused by the limited driving range and the considerable refueling time. Moreover, those who drive more (i.e., experience larger vehicle-miles of travel) are less interested in alternative fuel vehicles despite their more willingness-to-pay for electric driving range. Recently, Ref. 40 aims at eliciting the individuals' preferences for CV vs. BEV by estimating a hybrid choice model on a SP dataset collected in India. They find a willingness-to-pay for U.S.$7 to 40 to increase a kilometer in the driving range of a BEV with the current 200-kilometer range. Ref. 21 estimate a binary logit model to understand BEV adoption behavior of the middle-income working population in India, where they find acceptable driving range to be an influential factor.

## 2.2. Electric Driving Range: Psychological Aspect

The technological improvements in battery have led to longer driving range of the recently introduced BEVs in the 2020 vehicle market.[15,16] The issue of limited range can be further resolved through policy decision making geared towards improving publicly available PEV charging infrastructure.[5,17] The objective of a relevant study by Ref. 52 is reducing range anxiety through an effective design and placement of parking spaces and chargers to improve availability of charging infrastructure. To be effective, the results reveal the importance of the simultaneous implementation of these improvements.

In another line of research, a growing body of the literature exploring travel patterns of various cohorts of individuals conclude that the driving range of the existing affordable PEVs satisfy the required travel needs of a typical driver. For instance, in a work by Ref. 18 a simulation-based model is developed using a dataset selected from national travel surveys in Switzerland and Finland to study BEV driving range limitations. They find that BEVs available in the 2016 vehicle market can satisfy 85-90% of all national trip distances. This value can increase to 99% through adopting high-range BEVs and deploying public charging infrastructure policies. In another research example, Ref. 19 compare travel pattern of two groups including CV and BEV renters in the island Gotland, Sweden, wherein a network of charging infrastructure and EV rental schemes make the island "ready for EVs." A comparison of the driving pattern of the two groups through analyzing their route choice indicate that driving distance is not a limiting factor. In addition, they found that renting an EV instead of a CV leads to more knowledge and positive attitude towards EV.

Despite the above-mentioned driving range enhancements which are shown to be sufficient for most travel needs, recent studies report that the driving range factor still hinders PEV adoption (e.g., Ref. 21). This is likely because of the psychological anxiety caused by an irrational fear response unjustifiable by the driving range needed in reality,[23] i.e., range anxiety. This is also verified by the existing literature. For instance, the review study by Ref. 22 identifies driving range to be a psychological factor influencing consumer behavioral intention to, for instance, adopt, purchase, and use, which is concluded from reviewing studies relying on psychological theories and methods. A similar review study also categorizes the driving range factor as both technological and psychological factors.[13] Range anxiety itself is influenced by factors such as first-time experience.[53,54] For a review on studies focusing on BEV range satisfaction, anxiety, and utilization, and PHEV electric driving range, interested readers are referred to a systematic review by Ref. 55.

Relying on discrete choice models, Ref. 24 explore the impact of range concern, which is defined as an explanatory binary variable taking the value 1 for those who are concerned about PEV driving range and zero otherwise, on EV adoption behavior. For the empirical analysis, the authors estimate two binary logit models to explain the choice of PHEV and BEV versus other fuel types, and a multinomial probit model with a choice alternative set including CV, HEV, PHEV, and BEV. The study concludes that range anxiety is a top vehicle performance feature which needs to be addressed for promoting BEV mass adoption in the U.S.

### 2.3. Joint Decision on Vehicle Transaction Type and Electric Vehicle Adoption Behavior

The choice-based studies on EV adoption behavior analyze the decision process made by an analysis unit, e.g., an individual. In this process, given a set of choice alternatives including, for instance, CV, HEV, PHEV, and BEV, an individual chooses one as vehicle fuel type. This choice determines one dimension of the vehicle ownership problem (see Ref. 56 for a review) which can be simultaneously made with other vehicle dimensions such as vehicle transaction type. The rationale behind joint decision making is that an individual's decision on choosing, for instance, BEV as vehicle fuel type may be related to whether the BEV will be *traded for* an existing vehicle of the individual or will be *added* to his/her vehicle inventory.

A review of the relevant literature presented in Table 1 reveals that, despite offering valuable insights, there exists a limited number of studies on joint modeling of vehicle transaction and fuel type. An example is the study by Ref. 27 who estimate a mixed logit model to examine individuals' vehicle transaction type (including no-transaction, sell, and add) and fuel type (including CV, HEV, and BEV) in hypothetical scenarios designed for the next six years to mimic the vehicle market dynamics in the U.S. Another example is the study by Ref. 30 on individuals' choice from an alternative set encompassing retaining current vehicle, buying gasoline vehicle, buying HEV, and buying BEV. To consider random taste heterogeneity, they estimate a mixed logit model which also captures dynamics of HEV and BEV adoption behavior as reflected by a web-based database in a nine-year hypothetical time window in the U.S. While offering valuable insights, these studies do not consider comprehensive sets of vehicle transaction types and various EV types. The present study contributes to the literature by filling this gap in modeling the joint decision on vehicle fuel and transaction type.

To jointly estimate two vehicle decisions in a choice-based modeling framework, Ref. 57 and Ref. 58 suggest a multi-level NL model, which yields one vehicle decision at each level. Building on their framework, Ref. 49 examine Americans' vehicle transaction and fuel type decisions by estimating a three-level NL model, wherein the first level gives the vehicle transaction choice among no-transaction, sell, trade, and add. The second level determines fuel type of the added and traded-for vehicles between CV and PEV, and the third level gives the type of CV from the related alternatives including gasoline, diesel, and hybrid vehicles. They estimate the model using RP datasets that, however, is collected in two different areas due to the scarcity of RP dataset (though the results are verified based on spatial transferability criteria). Moreover, the datasets are collected in 2012-2013, which likely mimic the behavior of innovators and early adopters. Besides, their model does not distinguish between PHEV and BEV and bundles them as PEV. To have more consistent and reliable estimation results, the present study extends the study of Ref. 49 by modeling a NL model using a single dataset collected in 2019 (described in section 4) while also considering an inclusive alternative set and distinguishing PHEV and BEV adoption behavior.

## 3. Methodology

Figure 1 shows the framework of the two-level NL model with a latent construct explaining perceived range anxiety. The first level yields vehicle transaction type selected from the option set including making no transaction decision about a household vehicle (*no-transaction option*), selling an existing household vehicle (*sell option*), trading an existing household vehicle for another vehicle (*trade option*), and adding a vehicle to the existing household vehicle(s) (*add option*). The fuel type of the vehicles decided to be acquired, either as traded-for or added vehicles, is determined at the second level of the NL model simultaneously. The fuel type options include *CV*, *HEV*, *PHEV*, and *BEV*. Therefore, a vehicle transaction and fuel type choice set with ten alternatives is created, as shown in the left and middle sections of Figure 1. The model can distinctly characterize those who are interested in adding an EV type to the existing vehicle(s) of their household and those who are inclined to trade an existing vehicle for an EV type. This, together with distinguishing a comprehensive set of EV types (HEV, PHEV, and BEV), can provide more realistic empirical insights into the EV adoption behavior to center the policy recommendations on EV types that may replace CVs. Note that defining the choice alternatives at the vehicle level follows the approach of the prior studies on, for instance, vehicle transaction choice (e.g., Ref. 57 and Ref. 59), and vehicle transaction and fuel type choice (e.g., Ref. 30 and Ref. 49). Another approach, which is defining the choice alternatives at the household (or individual) level, raises the issue of multiple-choice observations for those who make multiple vehicle transaction choices, for instance, selling an existing vehicle and at the same time trading an existing vehicle for a BEV.

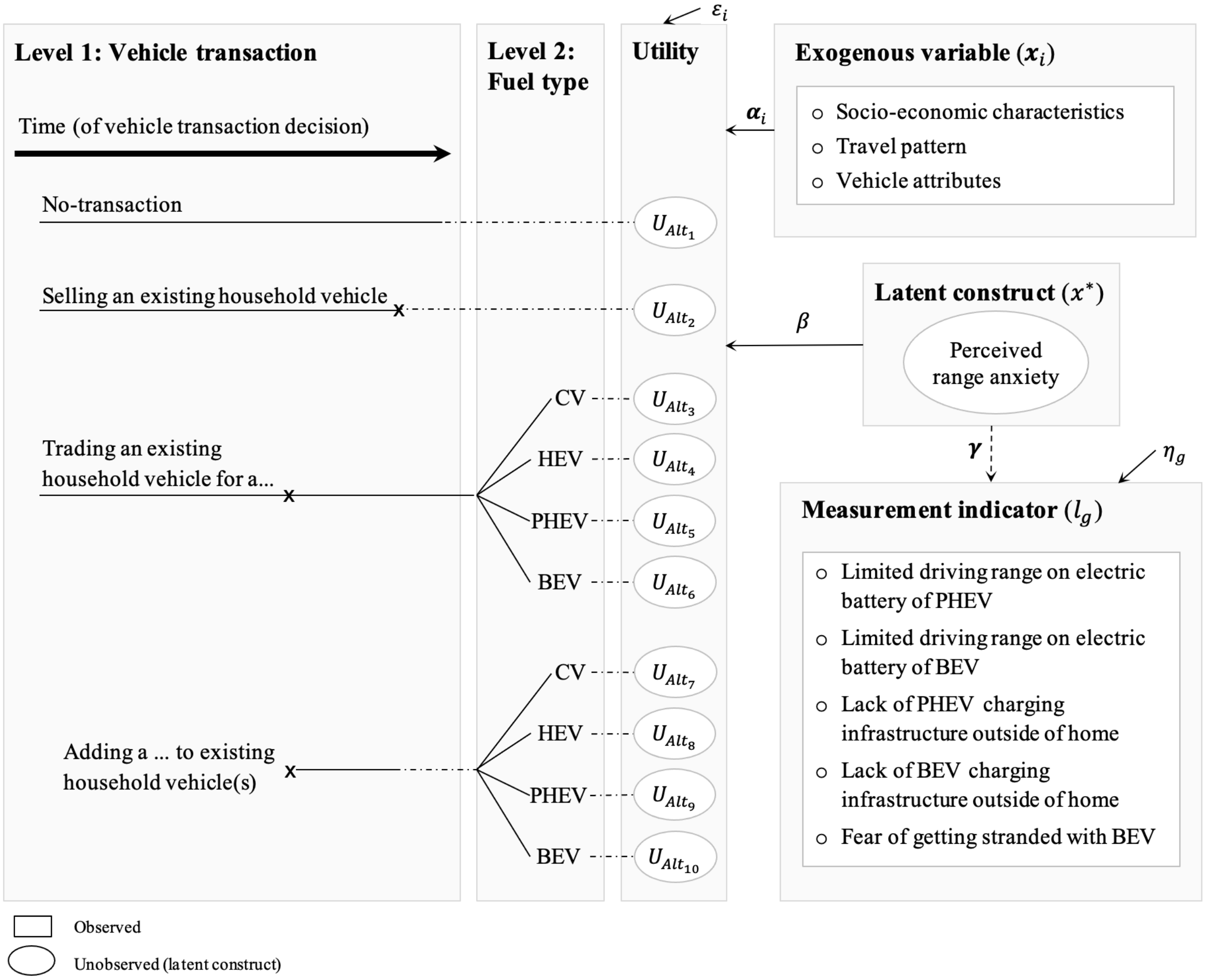

**Figure 1. The nested logit model of the joint vehicle transaction and fuel type choice (among ten alternatives) with a latent construct explaining perceived range anxiety**

The remainder of this section concisely presents the formulation of the well-known NL.[60] The NL model determines vehicle transaction and fuel type choice from the set of $I$ alternatives $[Alt_1, \ldots, Alt_I]$ $\forall i \in$ $\{1,2, \ldots, I = 10\}$, which are shown in the left side of Figure 1. For each observation of the sample dataset indexed as $q \in \{1,2, \ldots, Q = 3{,}536\}$, which is suppressed hereafter for brevity, an alternative is selected by the corresponding individual to maximize his/her utility. The utility of the $i^{\text{th}}$ alternative ($U_{Alt_i}$), which is expressed in equation (1), is as a function of the vector of exogenous variables ($\boldsymbol{x}_i$) and expected value of the latent construct ($\hat{x}^*$) via the corresponding coefficients denoted by vector $\boldsymbol{\alpha}_i$ and $\beta$, respectively. This equation is shown in Figure 1 by solid arrows connecting the latent construct and the vector of exogenous variables to the utility of choice alternatives. It is assumed that the random error component ($\varepsilon_i$) has extreme value distribution and there is no endogeneity bias (i.e., $E(\varepsilon_i|\boldsymbol{x}_i) = 0$ and $E(\varepsilon_i|\hat{x}^*) = 0$).

$$U_{Alt_i} = \boldsymbol{x}_i\boldsymbol{\alpha}_i + \hat{x}^*\beta + \varepsilon_i \qquad \forall i \in \{1,2, \ldots, I\} \tag{1}$$

The latent construct is determined by a CFA estimated on the sample dataset of 1,230 individual decision makers who are indexed as $n \in \{1,2, \ldots, N\}$. The index is suppressed hereafter for brevity. The CFA model is expressed by equation (2) which relates the latent construct ($x^*$) to the measurement indicators observed for the individuals ($l_g$) through the associated vector of loading factors ($\boldsymbol{\gamma}_g$). This relationship is shown by the dashed arrow connecting the latent construct to the measurement indicators in the bottom right of Figure 1. The effects of unobserved factors are denoted by the random error term ($\eta_g$) with a standard normal distribution — i.e., $\boldsymbol{\eta} \sim [\boldsymbol{0}, \boldsymbol{\Sigma}]$, where $\boldsymbol{\Sigma}$ is the covariance matrix.

$$l_g = x^*\boldsymbol{\gamma}_g + \eta_g \qquad \forall g \in \{1,2, \ldots, G\} \tag{2}$$

To write the likelihood function of the NL model, one needs to note that the model bundles $I$ alternatives in $J$ nests converting the choice set to $\left[\left(Alt_{1|1}, \ldots, Alt_{I_1|1}\right), \left(Alt_{1|2}, \ldots, Alt_{I_2|2}\right), \ldots, \left(Alt_{1|J}, \ldots, Alt_{I_J|J}\right)\right]$. In this study, $J$ equals the number of vehicle transaction types (i.e., no-transaction, sell, trade, and add). Consequently, the contribution of each vehicle transaction and fuel type choice observation to the likelihood function is the probability of choosing the $i^{\text{th}}$ alternative from the $j^{\text{th}}$ nest $\left(P_{ij}\right)$. Equation (3) explains this probability as the probability of selecting the $i^{\text{th}}$ alternative conditional on belonging to the $j^{\text{th}}$ nest $\left(P_{i|j}\right)$ multiplied by the marginal probability of the $j^{\text{th}}$ nest $\left(P_j\right)$, which are respectively calculated by equations (4) and (5). The marginal probability is determined by the inclusive value of the $j^{\text{th}}$ nest ($IV_j$), which is calculated as equation (6), through the corresponding parameter ($\gamma_j$). Then, the logarithm of the likelihood function over choice observations ($q \in \{1,2, \ldots, Q\}$), which is expressed by equation (7), is maximized using full information maximum likelihood to estimate the parameters.

$$P_{ij} = P_{i|j} \times P_j \tag{3}$$

$$P_{i|j} = \frac{\exp(\boldsymbol{x}_i \boldsymbol{\alpha}_i + \hat{x}^* \beta)}{\sum_{l=1}^{I_j} \exp(\boldsymbol{x}_i \boldsymbol{\alpha}_i + \hat{x}^* \beta)} \tag{4}$$

$$P_j = \frac{\exp[\gamma_j IV_j]}{\sum_{k=1}^{J} \exp[\gamma_k IV_k]} \tag{5}$$

$$IV_j = \ln\left(\sum_{i=1}^{I_j} \exp(\boldsymbol{x}_i \boldsymbol{\alpha}_i + \hat{x}^* \beta)\right) \tag{6}$$

$$\text{Log} - \text{likelihood} = \sum_{q=1}^{Q} \ln P_{ij,q} = \sum_{q=1}^{Q} \ln\left[P_{i|j,q} \times P_{j,q}\right] \tag{7}$$

# 4. Data

The NL model is empirically estimated on a sample dataset from the 2019 California Vehicle Survey conducted by California Energy Commission (2019) in the state of California.[61] The survey design and sampling plans follow the same patterns used in the previous waves of the data collection conducted in 2015-2017, which are thoroughly discussed in Ref. 62. The sample dataset selected for the present study contains 3,536 choice observations on the future vehicle transaction and fuel type decisions which correspond to SP of 1,230 households, each represented by an individual household member. The statistical analysis of the individual- and vehicle-related variables are respectively presented in sections 4.1 and 4.2.

## 4.1. Individual-Related Variables

Table 2 shows the statistical distribution of the sample individuals' socio-economic characteristics and travel pattern, which are the exogenous variables to the model. *The individuals' household structure* is explained by the number of adults living in the households, with an average value equal to 2.172, and the presence of child(ren), which is observed for 40.57% of the sample. *The individuals' household annual income* is categorized into three groups, including low (< \$75K), medium (\$75K ≤ < \$150K), and high (≥ \$150K) levels, which have the sample shares respectively at 39.43%, 39.02, and 21.55%. The distribution of *ethnicity* over the sample is so that 13.58% are Hispanic or Latino. *The residential type of individuals* is observed in four categories with the largest share belonging to single family detached house, which is observed for 70.16% of the sample. *The individuals' travel pattern* describes the use of ridesharing services, such as Uber and Lyft, which is observed for 24.31% of the individuals.

**Table 2. Statistical distribution of individual attributes (sample size = 1,230)**

| Exogenous variable | Category | # observations | Share (%) |
|---|---|---|---|
| *Socio-economic characteristic* | | | |
| Household structure | | | |
| Number of adults (age ≥ 18) | Mean = 2.172, SD = 0.856 | — | — |
| Presence of child(ren) | Yes | 499 | 40.57 |
| | No | 731 | 59.43 |
| Household annual income level | Low (< $75K) | 485 | 39.43 |
| | Medium ($75K ≤ < $150K) | 480 | 39.02 |
| | High (≥ $150K) | 265 | 21.55 |
| Ethnicity | Hispanic or Latino | 167 | 13.58 |
| | Non-Hispanic or Latino | 1,063 | 86.42 |
| Residential type | Single family detached house | 863 | 70.16 |
| | Single family attached house | 124 | 10.08 |
| | Apartment | 222 | 18.05 |
| | Other | 21 | 1.71 |
| *Travel pattern* | | | |
| Using ridesharing services | Yes | 299 | 24.31 |
| | No | 931 | 75.69 |

The individuals further respond to five questions measuring their *perceived concern about various aspects of electric driving range of PEVs*, which determine the latent perceived range anxiety construct. The responses are subsequently averaged over the sample individuals, as shown in Figure 2. Overall, it is observed that the concern about BEV-related measurement indicators is slightly higher than those of PHEV, which is consistent with the common sense. For instance, the average value of "lack of charging infrastructure outside of home" for BEV is larger than that of PHEV. It is worth mentioning that the indicator measuring "fear of getting stranded with BEV" is specific to BEV with a sample average value equal to 0.382.

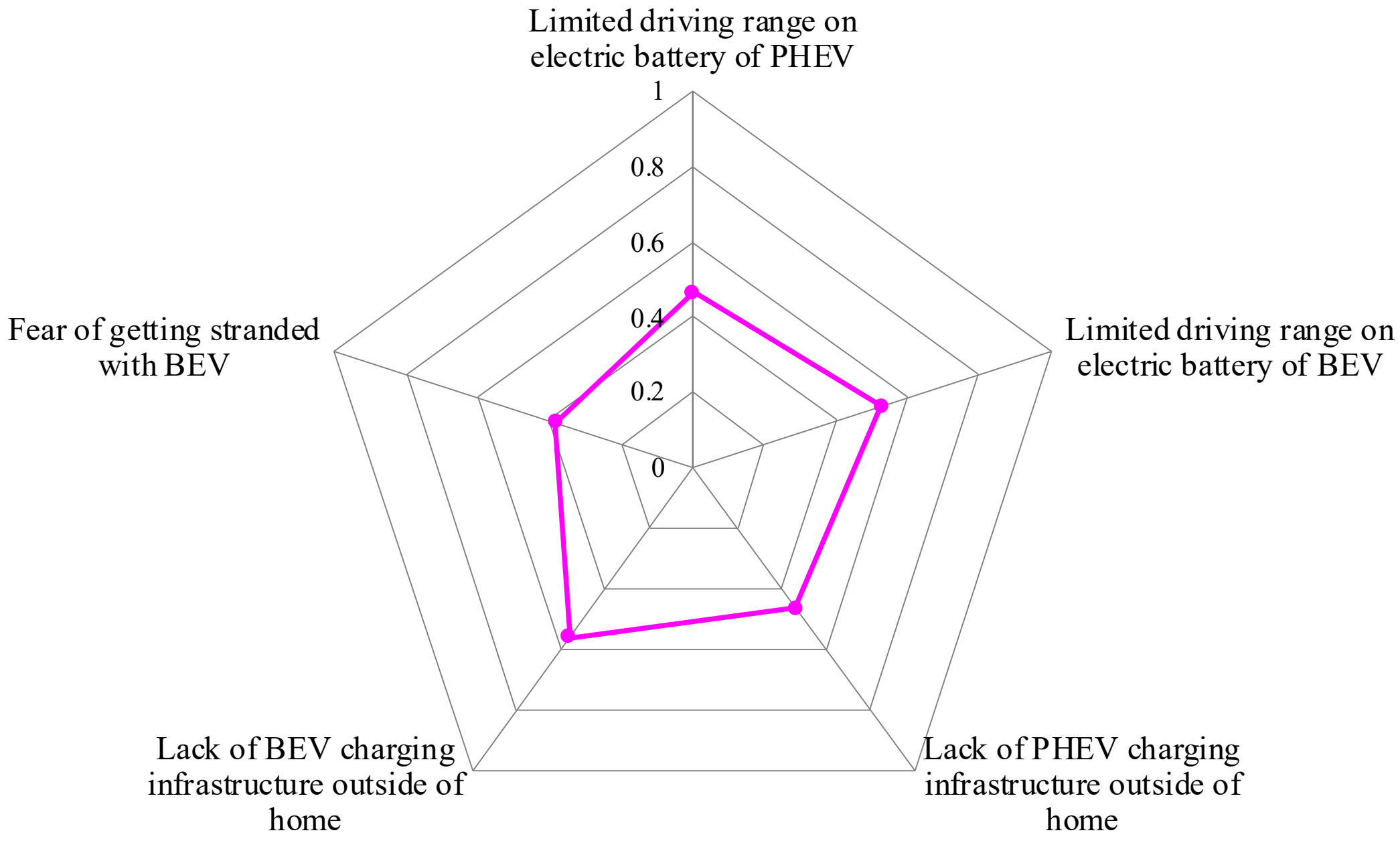

**Figure 2. The measurement indicators of plug-in electric vehicle range anxiety averaged over the sample individuals (sample size = 1,230)**

## 4.2. Vehicle-Related Variables

The statistical distribution of the vehicle-related variables is presented in Table 3. The first variable is the model outcome which describes *the choice of future vehicle transaction and fuel type decision* made by 1,230 households, each represented by an individual household member. Out of 3,536 choice observations, the shares of the choice set alternatives are shown in Table 3. The other two vehicle-related variables are exogenous to the choice model. *Vehicle age* before the transaction decision is observed to be on average 2.701 years with a standard deviation equal to 4.762 years. *Vehicle ownership* type before transaction has two types including owned and leased with the associated sample shares at 60.29% and 4.92%, respectively.

**Table 3. Statistical distribution of vehicle attributes (sample size = 3,536)**

| Variable | Category | # observations | Share (%) |
|---|---|---|---|
| *Outcome variable* | | | |
| Vehicle transaction and fuel types | No transaction — alternative 1 | 245 | 6.93 |
| | Sell — alternative 2 | 66 | 1.87 |
| | Trade, CV — alternative 3 | 1,247 | 35.27 |
| | Trade, HEV — alternative 4 | 499 | 14.11 |
| | Trade, PHEV — alternative 5 | 146 | 4.13 |
| | Trade, BEV — alternative 6 | 103 | 2.91 |
| | Add, CV — alternative 7 | 784 | 22.17 |
| | Add, HEV — alternative 8 | 292 | 8.26 |
| | Add, PHEV — alternative 9 | 83 | 2.35 |

| Variable | Category | # observations | Share (%) |
|---|---|---|---|
| | Add, BEV — alternative 10 | 71 | 2.01 |
| *Exogenous variable* | | | |
| Vehicle age before transaction (year) | Mean = 2.701; SD = 4.762<br>(Not applicable for add alternative) | — | — |
| Vehicle ownership type before transaction | Owned | 2,132 | 60.29 |
| | Leased | 174 | 4.92 |
| | Not applicable (i.e., add alternative) | 1,230 | 34.79 |

Note: CV = conventional gasoline and diesel vehicle; HEV = hybrid electric vehicle; PHEV = plug-in hybrid electric vehicle; BEV = battery electric vehicle.

# 5. Estimation Results

## 5.1. Confirmatory Factor Analysis

The latent construct explaining the individuals' perceived range anxiety is built on the five indicators, which measure the perceived concern about the various aspects of electric driving range of PEVs (see section 4.1 for the statistical analysis of the indicators). The weight (or loading factor) of the latent construct on each of the indicators is determined by estimating a CFA on the sample data for 1,230 individuals. The estimation results are presented in Table 4. To avoid the identification issue, the weights are normalized between -1 and 1, which are all found to be statistically significant at a 95% confidence interval.

**Table 4. Estimation results of confirmatory factor analysis**

| Indicators of range anxiety latent construct | Loading factor | t-stat |
|---|---|---|
| Limited driving range on electric battery of PHEV | 0.512 | 22.77 |
| Limited driving range on electric battery of BEV | 0.483 | 21.50 |
| Lack of PHEV charging infrastructure outside of home | 0.429 | 19.56 |
| Lack of BEV charging infrastructure outside of home | 0.427 | 19.45 |
| Fear of getting stranded with BEV | 0.234 | 10.77 |

Note: PHEV = plug-in hybrid electric vehicle; BEV = battery electric vehicle; Number of observations = 1,230; Goodness-of-fit index (GFI) = 0.992; Adjusted GFI = 0.958; Standardized root mean square residual (SRMR) = 0.033; Root mean square of error approximation (RMEA) = 0.008.

The estimated CFA has an acceptable goodness-of-fit concluded from the comparison of the goodness-of-fit measures with the corresponding critical values. Goodness-of-fit index (GFI) and adjusted GFI are respectively equal to 0.992 and 0.958, which are both larger than the critical value of 0.9 (suggested by Ref. 63). Standardized root mean square residuals (SRMR = 0.033) and root mean square of error approximation (RMEA = 0.008) meet the relevant criterion (<0.05), as suggested by Ref. 64 for SRMR, and Ref. 65 and Ref. 66 for RMEA. In addition, the model chi-square is statistically significant at a 95% confidence interval.[67] Looking at the estimated weights, the largest and the smallest weights belong to the indicators measuring "limited driving range on electric battery of PHEVs" and "fear of getting stranded with BEVs," respectively.

## 5.2. Nested Logit Model

The estimation results of the NL model of vehicle transaction and fuel type choice are presented in Table 5. Overall, the estimated model has an acceptable goodness-of-fit measured by rho-squared ($\rho^2 = 1 - LL(\beta) \,/\, LL(0)$) and McFadden's pseudo rho-squared ($\rho_c^2 = 1 - LL(\beta) \,/\, LL(c)$), which are respectively equal to 0.233 and 0.165. The nesting structure of the model, wherein level 1 yields the vehicle transaction type followed by the fuel type of the added and traded-for vehicles determined in level 2, is verified by the estimated inclusive value (IV) parameters of the add and the trade nests ranging between 0 and 1 at the 95% level of confidence (note that the IV parameters of the degenerated no-transaction and sell nests are fixed at 1.000). In other words, the nesting structure shows that the fuel type choice influences the choice of vehicle transaction type via the IV parameter values. Moreover, the estimated IV parameters of the add and the trade nests are closer to 1 than 0, indicating that vehicle transaction and fuel type choice are strongly interdependent, and thus the NL model better explains the choice than a multinomial choice model. The results are used to simulate the choice probability of the ten alternatives over the sample dataset. As shown in Figure 3, the simulated values (shown by the gray bars) almost match the corresponding observed market shares of the alternatives (shown by the pink bars).

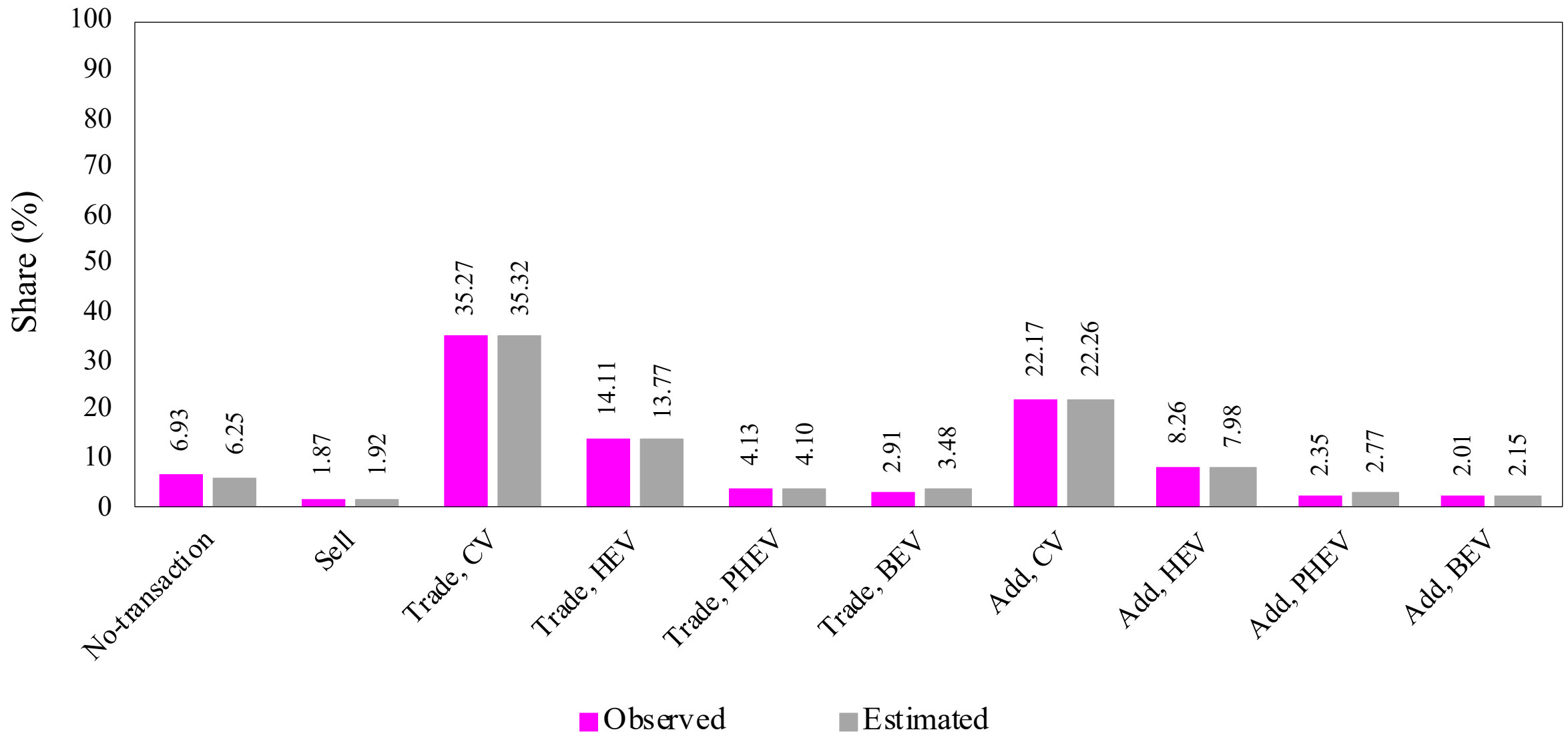

**Figure 3. Replication of sample dataset using the estimated nested logit model**

The results signify the *constant term* at a 95% confidence interval in the utility equations of the CV alternatives with a positive sign, and the sell alternative with a negative sign. It means that assuming all conditions are equal, CV, especially if added to the existing household vehicle(s), is more attractive than the other alternatives. In the same conditions, there is a possible inertia for selling an existing vehicle. The absolute values of the constant terms are acceptable given the absolute values of the other estimated coefficients multiplied by the corresponding variables. Furthermore, the constant term of the EV alternatives is found to be statistically insignificant implying that the individuals' preferences for these alternatives are almost explained by the explanatory variables, and not the "nominal" EV names. In other words, no influential factor on the utility of these alternatives is omitted from the model. The remainder of the section discusses the effects of the exogenous variables on the model outcome, which are all almost statistically significant at a 95% confidence interval.

### *5.2.1. Effects of Socio-Economic Characteristics*

*The household structure of the individuals* is explained by two variables; one is a count variable measuring the number of household adults who are aged 18 or higher. This variable is found to be negatively influencing the utility of the PEV-related alternatives, meaning that those who live with more adults likely are not interested in PEV neither as an added nor a traded-for vehicle. The other variable is dummy which takes the value 1 if a household has child(ren) and 0 otherwise. This variable enters the utility equation of the BEV-related alternatives, which shows the inertia caused by the presence of child(ren) towards acquiring BEV, either as an added or a traded-for vehicle. This finding is in line with a relevant review study in Ref. 13 and a prior empirical study in the State of California in Ref. 50. *The household annual income of the individuals* appears in the model as a dummy variable, which is equal to 1 for those with high income-level (i.e., $150K or higher). This group is found to be more interested in BEV, PHEV, and traded-for HEV in a descending manner. Given the higher ownership cost of EV, especially BEV, this behavior is consistent with the common sense and is also verified by the findings of a review in Ref. 47 and an empirical study in Ref. 49.

*Ethnicity* is found to be influential as a dummy variable equaling to 1 for Hispanic or Latino individuals. This variable appears in the utility equation of traded-for HEV, CV, and PHEV, where the magnitudes of the associated coefficients are respectively decreasing, revealing the tendency of this ethnic group towards these alternatives. *Residential type* enters the model as a dummy variable taking the value 1 for those who live in apartments. The results reveal the positive impact of this variable on HEV-related utility equations, revealing the positive tendency of the apartment residents towards HEV. The behavior of preferring HEV to PEV might be due to the fact that, compared with the single-family houses, apartments usually provide limited space required for installing charging infrastructure, while PEV users usually charge their vehicles at their residence as concluded by a review study by Ref. 68 and reported by Ref. 69. Moreover, a similar finding is reported by a relevant prior study in Ref. 49.

**Table 5. Estimation results of the nested logit model of vehicle transaction and fuel type choice**

| Exogenous variable | Sell | | Trade | | | | | | | | Add | | | | | | | |
|---|---|---|---|---|---|---|---|---|---|---|---|---|---|---|---|---|---|---|
| | | | CV | | HEV | | PHEV | | BEV | | CV | | HEV | | PHEV | | BEV | |
| | Coef. | t-stat | Coef. | t-stat | Coef. | t-stat | Coef. | t-stat | Coef. | t-stat | Coef. | t-stat | Coef. | t-stat | Coef. | t-stat | Coef. | t-stat |
| *Constant* | -1.695 | -10.05 | 1.325 | 17.06 | — | — | — | — | — | — | 1.320 | 16.10 | — | — | — | — | — | — |
| *Socio-economic characteristics* | | | | | | | | | | | | | | | | | | |
| Household structure | | | | | | | | | | | | | | | | | | |
| Number of adults (age ≥ 18) | — | — | — | — | — | — | -1.010 | -6.35 | -1.473 | -7.18 | — | — | — | — | -1.461 | -7.24 | -1.436 | -6.37 |
| With child(ren) (age < 18) | — | — | — | — | — | — | — | — | -0.283 | -1.96 | — | — | — | — | — | — | -0.457 | -1.93 |
| Household annual income level | | | | | | | | | | | | | | | | | | |
| High (≥ $150K) | — | — | — | — | 0.480 | 4.17 | 0.451 | 2.44 | 1.151 | 5.93 | — | — | — | — | 0.612 | 2.57 | 0.815 | 3.43 |
| Ethnicity | | | | | | | | | | | | | | | | | | |
| Hispanic or Latino | — | — | 1.473 | 7.77 | 1.495 | 6.68 | 1.093 | 3.36 | — | — | — | — | — | — | — | — | — | — |
| Residential type | | | | | | | | | | | | | | | | | | |
| Apartment | — | — | — | — | 0.457 | 3.43 | — | — | — | — | — | — | 0.621 | 4.03 | — | — | — | — |
| *Travel pattern* | | | | | | | | | | | | | | | | | | |
| Using ridesharing services | | | | | | | | | | | | | | | | | | |
| Yes | — | — | — | — | 0.521 | 4.65 | — | — | — | — | — | — | 0.548 | 3.93 | — | — | — | — |
| *Vehicle attributes* | | | | | | | | | | | | | | | | | | |
| Logarithm of vehicle age (year) | 0.888 | 7.42 | 0.717 | 9.99 | 0.648 | 7.39 | 0.412 | 3.26 | 0.299 | 1.96 | — | — | — | — | — | — | — | — |
| Vehicle ownership type | | | | | | | | | | | | | | | | | | |
| Leased | — | — | 2.749 | 6.63 | 3.185 | 7.33 | 3.698 | 8.13 | 3.454 | 7.12 | — | — | — | — | — | — | — | — |
| *Latent construct* | | | | | | | | | | | | | | | | | | |
| Perceived range anxiety | — | — | — | — | — | — | — | — | -1.054 | -2.02 | — | — | — | — | — | — | -1.302 | -2.09 |
| *Inclusive value parameter* | 1.000 | — | 0.943 | 18.94 | | | | | | | 0.926 | 17.25 | | | | | | |

Note: CV = conventional gasoline and diesel vehicle; HEV = hybrid electric vehicle; PHEV = plug-in hybrid electric vehicle; BEV = battery electric vehicle; The base alternative is no-transaction; Number of observations = 3,536; Log-likelihood at convergence $\left(LL(\beta)\right)$ = -6,246.551; Log-likelihood at zero $\left(LL(0)\right)$ = -8,141.941; Log-likelihood at constants $\left(LL(c)\right)$ = -7,484.911.

### *5.2.2. Effects of Travel Pattern*

*The individuals' travel pattern* is described by a dummy variable taking the value 1 if an individual uses ridesharing services such as Uber and Lyft. The results, which signify this variable in the utility equation of the HEV-related alternatives with a positive sign, reveal the preference of the ridesharing users for HEV with a slightly higher tendency towards added than traded-for HEV, as indicated by the slightly larger relevant coefficient. This implies that policies targeting the reduction of vehicle ownership through encouraging ridesharing use may not be effective, as ridesharing use may lead to a slightly higher vehicle ownership level, which is inferred from the slightly higher tendency of the ridesharing users towards adding a HEV to their existing vehicle(s) than trading an existing one for a HEV.

### *5.2.3. Effects of Vehicle Attributes*

*Vehicle age* enters the model in a logarithmic form with a positive sign in the utility equations of the sell and all four trade alternatives. This suggests that as the age of a vehicle increases, the probability of being sold and traded for CV, HEV, PHEV, and BEV increases in a descending order. This is consistent with the common sense as well as the findings of the prior studies in Ref. 49. The results furthermore reveal the logarithmic form of this variable in the model implying the associated diminishing impact for older vehicles.

*Vehicle ownership type* appears in the model as a dummy variable equaling 1 for leased vehicles. The appearance of this variable in the utility equations of all four traded-for fuel types indicates that individuals who lease a vehicle most likely trade in a vehicle. Also, their tendency in a descending order is towards PHEV, BEV, HEV, and CV, as indicated by the magnitudes of the corresponding coefficients. This finding is consistent with the prior studies which report the positive impact of leasing a vehicle on BEV adoption in the State of California (see Ref. 50) and Gotland, Sweden (see Ref. 19) as well as on trading a vehicle for HEV, PHEV, or BEV in the U.S. (see Ref. 51). The reason for observing such behavior might be the differences between buying and leasing a vehicle. In fact, the interest of persons who lease a vehicle in the three EV types implies EV attractiveness to attentive persons to factors such as access to late-model vehicles, coverage of the vehicle's maintenance costs, and no concern about the drop in the vehicle's trade-in value, which can be listed as the advantages of leasing a vehicle. Moreover, vehicles that are leased usually have the limitation of miles driven monthly. This implies that persons whose mobility needs are met by less vehicle use, in terms of vehicle-miles of travel (VMT), are likely more interested in the three EV types compared to the persons with more vehicle use, i.e., larger VMT.

### *5.2.4. Effects of Latent Construct*

After trying the latent construct in the model, we found this variable only influencing the utility of the BEV-related alternatives with an intuitive negative sign. This means that individuals who perceive higher anxiety about driving range will less likely select BEV either as an added or a traded-for vehicle. Moreover, the statistically insignificance of this factor in the PHEV-related options means that individuals' choice of PHEV cannot be associated with perceived range anxiety. This finding is in line with the prior studies. Ref.

24 recognizes range anxiety as an essential performance component to promote BEV in the U.S., as concluded from the significance of the factor of vehicle range concern in choosing BEV and not PHEV. Ref. 42 finds that battery range only affects BEV but not PHEV.

Comparing the two coefficients in the utility equations of BEVs further reveals that the negative tendency towards adding is more than trading BEV. Recalling the definition of the latent constructs built on the corresponding indicators (as discussed in section 5.1 on the estimation results of CFA), we can conclude that persons who are more anxious about the limited driving range and lack of charging infrastructure of PHEV and BEV, and getting stranded with BEV less likely opt for BEV with a lower tendency towards adding than trading in BEV.

## 6. Concluding Remarks

Electric vehicles (EVs) can replace fossil-fueled conventional vehicles (CVs) in the pathway to net zero emissions. Yet, a major barrier to the widespread adoption of various EV types, especially plug-in EV (PEV), is limited electric driving range. Despite the *technical* improvements in driving range through advances in battery technology and policies on improving charging infrastructure, driving range is still reported as an obstacle to EV adoption. Thus, the focus of the research on EV adoption behavior needs to shift to the *psychological* aspect of driving range which is referred to as "range anxiety" capturing the psychological anxiety caused by an irrational fear response unjustifiable by the driving range needed in reality. With this aim, this study presents a nested logit (NL) model with a latent construct explaining perceived anxiety about electric driving range. The model captures the adoption behavior of various EV types including hybrid EV (HEV), plug-in HEV (PHEV), and battery EV (BEV) while also emphasizing on vehicle transaction type. The empirical results of the model on a sample dataset from the State of California, U.S. are utilized to characterize the potential adopters of various EV types distinguished by vehicle transaction type. Notably, persons with higher perceived anxiety will less likely adopt BEV, especially as an added than a traded-for vehicle, while this factor does not significantly change their preference for PHEV. The interest in PEV, especially BEV, is attributed to high household income level (i.e., $150K or higher) and the presence of not many adult household members. Those who live with child(ren) probably do not adopt BEV with less intention towards added than traded-for BEVs, while this characteristic does not significantly influence PHEV adoption. The remarkable attribute of HEV adopters probably is the use of ridesharing services. Moreover, if a vehicle is leased by a person, especially vehicles with higher age, the likelihood of trading it with a PHEV and a BEV is high in a descending order.

To relax the modeling and sample data limitations, future research areas are suggested in three directions. The first direction is extending the research scope by jointly modeling multiple vehicle-related decisions, such as transaction, fuel, and use (in terms of vehicle-miles of travel), or incorporating the time dimension into transaction and fuel decisions, which can be approached by joint discrete-continuous frameworks. The second direction of studies can improve the limitations of the dataset used in this study, which was acquired from a survey asking respondents about the features of their expected future vehicles, but did not constitute a full-fledged stated choice experiment in a controlled setting with detailed vehicle attributes such as purchase price and fuel cost. Lastly, although the survey used for the empirical analysis of this research is one of the relatively recent available ones conducted in 2019, people's opinions and preferences about EVs might have changed and EVs have penetrated more into markets since then. Therefore, further research can conduct new surveys and employ the modeling framework suggested in this study to compare the (dis)similarities in empirical findings over time (an example is a dynamic model of vehicle transaction and fuel type choice using a retrospective vehicle survey in the U.S. as in Ref. 51).

## Acknowledgement

This work was funded in part by the U.S. National Science Foundation (Award No. 2112650). The authors are solely responsible for the findings of this research. The opinions expressed are solely those of the authors, and do not necessarily represent those of the US National Science Foundation. The authors are grateful to the California Energy Commission for database provision.

## Author Declarations

### Conflict of Interest

The authors have no conflicts to disclose.

### Author Contributions

**Fatemeh Nazari:** Conceptualization; Formal analysis; Methodology; Data curation; Software; Writing – original draft. **Abolfazl (Kouros) Mohammadian:** Conceptualization; Methodology; Supervision; Writing – review & editing.

## Data Availability

The data that support the findings of this study are openly available in California Energy Commission at https://www.energy.ca.gov/data-reports/surveys/california-vehicle-survey. Ref. 61

## References

[1]G. Santos, H. Behrendt, L. Maconi, T. Shirvani, and A. Teytelboym, "Part I: Externalities and economic policies in road transport," Research in Transportation Economics 28(1), 2–45 (2010). https://doi.org/10.1016/j.retrec.2009.11.002

[2]European Environment Agency (EEA), Health Impacts of Exposure to Noise from Transport in Europe. Retrieved from: https://www.eea.europa.eu/en/analysis/indicators/health-impacts-of-exposure-to-1 (2025).

[3]International Energy Agency (IEA), Transport: Improving the Sustainability of Passenger and Freight Transport. Retrieved from: https://www.iea.org/energy-system/transport (IEA, Paris, 2023).

[4]C. Liu, and Z. Lin, "How uncertain is the future of electric vehicle market: Results from Monte Carlo simulations using a nested logit model," International Journal of Sustainable Transportation 11(4), 237–247 (2017). https://doi.org/10.1080/15568318.2016.1248583

[5]International Energy Agency (IEA), Global EV Outlook 2024: Securing Supplies for an Electric Future. Retrieved from: https://www.iea.org/reports/global-ev-outlook-2024 (2024).

[6]Alliance for Automotive Innovation, U.S. Light-Duty Advanced Technology Vehicle (ATV) Sales (2011-2025). Retrieved from: https://www.autosinnovate.org/EVDashboard (2025).

[7]H.S. Boudet, "Public perceptions of and responses to new energy technologies," Nature Energy 4(6), 446–455 (2019). https://doi.org/10.1038/s41560-019-0399-x

[8]E. Costa, A. Horta, A. Correia, J. Seixas, G. Costa, and D. Sperling, "Diffusion of electric vehicles in Brazil from the stakeholders' perspective," International Journal of Sustainable Transportation 15(11), 865–878 (2021). https://doi.org/10.1080/15568318.2020.1827317

[9]M.M. Hamed, A. Mustafa, M. Al-Sharif, and M. Shawaqfah, "Modeling the households' satisfaction level with the first electric vehicle and the time until the purchase of the second electric vehicle," International Journal of Sustainable Transportation 17(1), 52–64 (2023). https://doi.org/10.1080/15568318.2021.1983677

[10]O. Egbue, and S. Long, "Barriers to widespread adoption of electric vehicles: An analysis of consumer attitudes and perceptions," Energy Policy 48, 717–729 (2012). https://doi.org/10.1016/j.enpol.2012.06.009

[11]A. Hackbarth, and R. Madlener, "Willingness-to-pay for alternative fuel vehicle characteristics: A stated choice study for Germany," Transportation Research Part A: Policy and Practice 85, 89–111 (2016). https://doi.org/10.1016/j.tra.2015.12.005

[12]J. Kester, G.Z. de Rubens, B.K. Sovacool, and L. Noel, "Public perceptions of electric vehicles and vehicle-to-grid (V2G): Insights from a Nordic focus group study," Transportation Research Part D: Transport and Environment 74, 277–293 (2019). https://doi.org/10.1016/j.trd.2019.08.006

[13]V. Singh, V. Singh, and S. Vaibhav, "A review and simple meta-analysis of factors influencing adoption of electric vehicles," Transportation Research Part D: Transport and Environment 86, 102436 (2020). https://doi.org/10.1016/j.trd.2020.102436

[14]E. Higueras-Castillo, A. Guillén, L.-J. Herrera, and F. Liébana-Cabanillas, "Adoption of electric vehicles: Which factors are really important?" International Journal of Sustainable Transportation 15(10), 799–813 (2021). https://doi.org/10.1080/15568318.2020.1818330

[15]J.A. Sanguesa, V. Torres-Sanz, P. Garrido, F.J. Martinez, and J.M. Marquez-Barja, "A review on electric vehicles: Technologies and challenges," Smart Cities 4(1), 372–404 (2021). https://doi.org/10.3390/smartcities4010022

[16]U.S. Environmental Protection Agency (US EPA), Office of Energy Efficiency and Renewable Energy, Fuel Economy Guide, Model Year 2020 (2023).

[17]The White House, Fact Sheet: Biden-Harris Administration Announces New Standards and Major Progress for a Made-in-America National Network of Electric Vehicle Chargers. Retrieved from: https://www.whitehouse.gov/Briefing-Room/Statements-Releases/2023/02/15/Fact-Sheet-Biden-Harris-Administration-Announces-New-Standards-and-Major-Progress-for-a-Made-in-America-National-Network-of-Electric-Vehicle-Chargers/ (2023).

[18]M.A. Melliger, O.P. van Vliet, and H. Liimatainen, "Anxiety vs reality – Sufficiency of battery electric vehicle range in Switzerland and Finland," Transportation Research Part D: Transport and Environment 65, 101–115 (2018). https://doi.org/10.1016/j.trd.2018.08.011

[19]J.H. Langbroek, M. Cebecauer, J. Malmsten, J.P. Franklin, Y.O. Susilo, and P. Georén, "Electric vehicle rental and electric vehicle adoption," Research in Transportation Economics 73, 72–82 (2019). https://doi.org/10.1016/j.retrec.2019.02.002

[20]G. Brancaccio, and F.P. Deflorio, "Extracting travel patterns from floating car data to identify electric mobility needs: A case study in a metropolitan area," International Journal of Sustainable Transportation 17(2), 181–197 (2023). https://doi.org/10.1080/15568318.2021.2004629

[21]T. Munshi, S. Dhar, and J. Painuly, "Understanding barriers to electric vehicle adoption for personal mobility: A case study of middle income in-service residents in Hyderabad city, India," Energy Policy 167, 112956 (2022). https://doi.org/10.1016/j.enpol.2022.112956

[22]W. Li, R. Long, H. Chen, and J. Geng, "A review of factors influencing consumer intentions to adopt battery electric vehicles," Renewable and Sustainable Energy Reviews 78, 318–328 (2017). https://doi.org/10.1016/j.rser.2017.04.076

[23]J.B. Griffin, Clinical Methods: The History, Physical, and Laboratory Examinations, Anxiety (Chapter 202). Retrieved from: https://www.ncbi.nlm.nih.gov/books/NBK315/ (Butterworths, Boston, U.S., 1990).

[24]B.W. Lane, J. Dumortier, S. Carley, S. Siddiki, K. Clark-Sutton, and J.D. Graham, "All plug-in electric vehicles are not the same: Predictors of preference for a plug-in hybrid versus a battery-electric vehicle,"

Transportation Research Part D: Transport and Environment 65, 1–13 (2018). https://doi.org/10.1016/j.trd.2018.07.019

[25]S. Beggs, S. Cardell, and J. Hausman, "Assessing the potential demand for electric cars," Journal of Econometrics 17(1), 1–19 (1981). https://doi.org/10.1016/0304-4076(81)90056-7

[26]D. Brownstone, D.S. Bunch, T.F. Golob, and W. Ren, "A transactions choice model for forecasting demand for alternative-fuel vehicles," Research in Transportation Economics 4, 87–129 (1996). https://doi.org/10.1016/S0739-8859(96)80007-2

[27]M. Maness, and C. Cirillo, "Measuring future vehicle preferences: Stated preference survey approach with dynamic attributes and multiyear time frame," Transportation Research Record: Journal of the Transportation Research Board 2285(1), 100–109 (2012). https://doi.org/10.3141/2285-12

[28]A. Hoen, and M.J. Koetse, "A choice experiment on alternative fuel vehicle preferences of private car owners in the Netherlands," Transportation Research Part A: Policy and Practice 61, 199–215 (2014). https://doi.org/10.1016/j.tra.2014.01.008

[29]B. Smith, D. Olaru, F. Jabeen, and S. Greaves, "Electric vehicles adoption: Environmental enthusiast bias in discrete choice models," Transportation Research Part D: Transport and Environment 51, 290–303 (2017). http://dx.doi.org/10.1016/j.trd.2017.01.008

[30]C. Cirillo, Y. Liu, and M. Maness, "A time-dependent stated preference approach to measuring vehicle type preferences and market elasticity of conventional and green vehicles," Transportation Research Part A: Policy and Practice 100, 294–310 (2017). https://doi.org/10.1016/j.tra.2017.04.028

[31]F. Liao, E. Molin, H. Timmermans, and B. van Wee, "The impact of business models on electric vehicle adoption: A latent transition analysis approach," Transportation Research Part A: Policy and Practice 116, 531–546 (2018).

[32]R. Wolbertus, M. Kroesen, R. van den Hoed, and C.G. Chorus, "Policy effects on charging behaviour of electric vehicle owners and on purchase intentions of prospective owners: Natural and stated choice experiments," Transportation Research Part D: Transport and Environment 62, 283–297 (2018). https://doi.org/10.1016/j.trd.2018.03.012

[33]Y. Kwon, S. Son, and K. Jang, "Evaluation of incentive policies for electric vehicles: An experimental study on Jeju Island," Transportation Research Part A: Policy and Practice 116, 404–412 (2018). https://doi.org/10.1016/j.tra.2018.06.015

[34]B. Leard, "Consumer inattention and the demand for vehicle fuel cost savings," Journal of Choice Modelling 29, 1–16 (2018). https://doi.org/10.1016/j.jocm.2018.08.002

[35]F. Liao, E. Molin, H. Timmermans, and B. van Wee, "Consumer preferences for business models in electric vehicle adoption," Transport Policy 73, 12–24 (2019). https://doi.org/10.1016/j.tranpol.2018.10.006

[36]L. Qian, J.M. Grisolía, and D. Soopramanien, "The impact of service and government-policy attributes on consumer preferences for electric vehicles in China," Transportation Research Part A: Policy and Practice 122, 70–84 (2019). https://doi.org/10.1016/j.tra.2019.02.008

[37]M. Ghasri, A. Ardeshiri, and T. Rashidi, "Perception towards electric vehicles and the impact on consumers' preference," Transportation Research Part D: Transport and Environment 77, 271–291 (2019). https://doi.org/10.1016/j.trd.2019.11.003

[38]L. Li, Z. Wang, L. Chen, and Z. Wang, "Consumer preferences for battery electric vehicles: A choice experimental survey in China," Transportation Research Part D: Transport and Environment 78, 102185 (2020). https://doi.org/10.1016/j.trd.2019.11.014

[39]E. Guerra, and R.A. Daziano, "Electric vehicles and residential parking in an urban environment: Results from a stated preference experiment," Transportation Research Part D: Transport and Environment 79, 102222 (2020). https://doi.org/10.1016/j.trd.2020.102222

[40]P. Bansal, R.R. Kumar, A. Raj, S. Dubey, and D.J. Graham, "Willingness to pay and attitudinal preferences of Indian consumers for electric vehicles," Energy Economics 100, 105340 (2021). https://doi.org/10.1016/j.eneco.2021.105340

[41]S. Khan, H. Maoh, and T. Dimatulac, "The demand for electrification in Canadian fleets: A latent class modeling approach," Transportation Research Part D: Transport and Environment 90, 102653 (2021). https://doi.org/10.1016/j.trd.2020.102653

[42]W. Jia, and T.D. Chen, "Are Individuals' stated preferences for electric vehicles (EVs) consistent with real-world EV ownership patterns?" Transportation Research Part D: Transport and Environment 93, 102728 (2021). https://doi.org/10.1016/j.trd.2021.102728

[43]Z. Rezvani, J. Jansson, and J. Bodin, "Advances in consumer electric vehicle adoption research: A review and research agenda," Transportation Research Part D: Transport and Environment 34, 122–136 (2015). https://doi.org/10.1016/j.trd.2014.10.010

[44]S. Haustein, and A.F. Jensen, "Factors of electric vehicle adoption: A comparison of conventional and electric car users based on an extended theory of planned behavior," International Journal of Sustainable Transportation 12(7), 484–496 (2018). https://doi.org/10.1080/15568318.2017.1398790

[45]S.-J. Moon, "Effect of consumer environmental propensity and innovative propensity on intention to purchase electric vehicles: Applying an extended theory of planned behavior," International Journal of Sustainable Transportation 16(11), 1032–1046 (2021). https://doi.org/10.1080/15568318.2021.1961950

[46]M. Pei, Z. Huang, Z. Zhang, K. Wang, and X. Ye, "Range anxiety and willingness to pay: Psychological insights for electric vehicle," Journal of Renewable and Sustainable Energy 17(1), (2025). https://doi.org/10.1063/5.0220237

[47]F. Liao, E. Molin, and B. van Wee, "Consumer preferences for electric vehicles: A literature review," Transport Reviews 37(3), 252–275 (2017). https://doi.org/10.1080/01441647.2016.1230794

[48]R.J. Javid, and A. Nejat, "A comprehensive model of regional electric vehicle adoption and penetration," Transport Policy 54, 30–42 (2017). https://doi.org/10.1016/j.tranpol.2016.11.003

[49]F. Nazari, A.K. Mohammadian, and T. Stephens, "Dynamic household vehicle decision modeling considering plug-In electric vehicles," Transportation Research Record: Journal of the Transportation Research Board, 1–10 (2018). https://doi.org/10.1177/0361198118796925

[50]F. Nazari, A.K. Mohammadian, and T. Stephens, "Modeling electric vehicle adoption considering a latent travel pattern construct and charging infrastructure," Transportation Research Part D: Transport and Environment 72, 65–82 (2019). https://doi.org/10.1016/j.trd.2019.04.010

[51]F. Nazari, M. Noruzoliaee, and A. (Kouros) Mohammadian, "Electric vehicle adoption behavior and vehicle transaction decision: Estimating an integrated choice model with latent variables on a retrospective vehicle survey," Transportation Research Record: Journal of the Transportation Research Board, 1–20 (2023). https://doi.org/10.1177/03611981231184875

[52]H.A. Bonges III, and A.C. Lusk, "Addressing electric vehicle (EV) sales and range anxiety through parking layout, policy and regulation," Transportation Research Part A: Policy and Practice 83, 63–73 (2016). http://dx.doi.org/10.1016/j.tra.2015.09.011

[53]T. Franke, N. Rauh, M. Günther, M. Trantow, and J.F. Krems, "Which factors can protect against range stress in everyday usage of battery electric vehicles? Toward enhancing sustainability of electric mobility systems," Human Factors 58(1), 13–26 (2016). https://doi.org/10.1177/0018720815614702

[54]N. Rauh, T. Franke, and J.F. Krems, "First-time experience of critical range situations in BEV use and the positive effect of coping information," Transportation Research Part F: Traffic Psychology and Behaviour 44, 30–41 (2017). http://dx.doi.org/10.1016/j.trf.2016.10.001

[55]E. Daramy-Williams, J. Anable, and S. Grant-Muller, "A systematic review of the evidence on plug-in electric vehicle user experience," Transportation Research Part D: Transport and Environment 71, 22–36 (2019). https://doi.org/10.1016/j.trd.2019.01.008

[56]S. Anowar, N. Eluru, and L.F. Miranda-Moreno, "Alternative modeling approaches used for examining automobile ownership: A comprehensive review," Transport Reviews 34(4), 441–473 (2014). https://doi.org/10.1080/01441647.2014.915440

[57]A. (Kouros) Mohammadian, and E. Miller, "Dynamic modeling of household automobile transactions," Transportation Research Record: Journal of the Transportation Research Board 1831(1), 98–105 (2003). https://doi.org/10.3141/1831-11

[58]A. (Kouros) Mohammadian, and E. Miller, "Empirical investigation of household vehicle type choice decisions," Transportation Research Record: Journal of the Transportation Research Board (1854), 99–106 (2003). https://doi.org/10.3141/1854-11

[59]M.J. Roorda, J.A. Carrasco, and E.J. Miller, "An integrated model of vehicle transactions, activity scheduling, and mode choice," Transportation Research Part B: Methodological 43(2), 217–229 (2009). https://doi.org/10.1016/j.trb.2008.05.003

[60]H.C. Williams, "On the formation of travel demand models and economic evaluation measures of user benefit," Environment and Planning A 9(3), 285–344 (1977). https://doi.org/10.1068/a090285

[61]California Energy Commission, 2019 California Vehicle Survey. Retrieved from: https://www.energy.ca.gov/data-reports/surveys/california-vehicle-survey (2019).

[62]M. Fowler, T. Cherry, T. Adler, M. Bradley, and A. Richard, 2015-2017 California Vehicle Survey, Prepared for California Energy Commission. Retrieved from: https://www.energy.ca.gov/sites/default/files/2023-02/CEC-200-2018-006.pdf (2018).

[63]Y. Gao, S. Rasouli, H. Timmermans, and Y. Wang, "Understanding the relationship between travel satisfaction and subjective well-being considering the role of personality traits: A structural equation model," Transportation Research Part F: Traffic Psychology and Behaviour 49, 110–123 (2017). https://doi.org/10.1016/j.trf.2017.06.005

[64]B.M. Byrne, Structural Equation Modeling with AMOS: Basic Concepts, Applications, and Programming, 3rd Edition (Routledge, New York, NY, USA, 2016). https://doi.org/10.4324/9781315757421

[65]J.H. Steiger, "Structural model evaluation and modification: An interval estimation approach," Multivariate Behavioral Research 25(2), 173–180 (1990). https://doi.org/10.1207/s15327906mbr2502_4

[66]M.W. Browne, and R. Cudeck, "Alternative ways of assessing model fit," Sociological Methods & Research 21(2), 230–258 (1992). https://doi.org/10.1177/0049124192021002005

[67]T.F. Golob, "Structural equation modeling for travel behavior research," Transportation Research Part B: Methodological 37(1), 1–25 (2003). https://doi.org/10.1016/S0191-2615(01)00046-7

[68]S. Hardman, A. Jenn, G. Tal, J. Axsen, G. Beard, N. Daina, E. Figenbaum, N. Jakobsson, P. Jochem, and N. Kinnear, "A review of consumer preferences of and interactions with electric vehicle charging infrastructure," Transportation Research Part D: Transport and Environment 62, 508–523 (2018). https://doi.org/10.1016/j.trd.2018.04.002

[69]U.S. Department of Energy (US DOE), Office of Energy Efficiency & Renewable Energy, Charging Electric Vehicles at Home. Retrieved from: https://afdc.energy.gov/fuels/electricity-charging-home (2023).